\documentstyle{article}
\newtheorem{theorem}{Theorem}
\newtheorem{lmm}[theorem]{Lemma}

\newtheorem{pro}[theorem]{Proposition}
\newtheorem{df}[theorem]{Definition}

\newtheorem{ass}[theorem]{Assumption}

\newcommand\calA{{\cal A}}

\newcommand\calF{{\cal F}}

\newcommand\calH{{\cal H}}

\newcommand\calK{{\cal K}}

\newcommand\calM{{\cal M}}
\newcommand\calN{{\cal N}}

\newcommand\calS{{\cal S}}

\newcommand\bfZ{\bf Z}

\newcommand\hal{H_\Lambda}

\newcommand\evl{\alpha_t}

\newcommand{\abs}[1]{\left|#1\right|}
\newcommand{\norm}[1]{\left\Vert#1\right\Vert}

\newcommand{\rbk}[1]{\left(#1\right)}
\newcommand{\sbk}[1]{\left[#1\right]}
\newcommand{\cbk}[1]{\left\{#1\right\}}
\newcommand{\tri}[1]{\Vert\vert#1\vert\Vert}

\begin{document}
\newpage\thispagestyle{empty}
{\topskip 2cm
\begin{center}
{\Huge\bf
On the Algebra of
Fluctuation in Quantum Spin Chains.
  }
\\
\bigskip\bigskip
\bigskip\bigskip
\bigskip\bigskip
{\Large Taku Matsui}
\\
\bigskip
{\it Graduate School of Mathematics
\\
Kyushu University
\\
1-10-6 Hakozaki, Fukuoka 812-8581
\\
JAPAN}
\end{center}
\vfil
\noindent
We present a proof of the central limit theorem
for a pair of mutually non-commuting operators
in mixing quantum spin chains.
The operators are not necessarily strictly local
but quasi-local. As a corollary we
obtain a direct construction of the time evolution of 
the algebra of normal fluctuation 
for Gibbs states of finite range interactions
on a one-dimensional lattice.
We show that the state of the algebra of normal fluctuation
satisfies the $\beta$-KMS condition if the microscopic state is 
a $\beta$-KMS state.
\par
We show that any mixing finitely correlated state satisfies
our assumption for the central limit theorem.
\noindent
\bigskip
\hrule
\noindent
{\small\tt e-mail: matsui@math.kyushu-u.ac.jp}
\vfil}\newpage
\section{Summary of Results}
\setcounter{theorem}{0}
\setcounter{equation}{0}
In \cite{Goderis1}  and in \cite{Goderis2}
 D.Goderis, A.Verbeure, and P.Vets investigated
a central limit theorem for  mixing quantum spin systems.
 We refer to the limit theorem of  D.Goderis, A.Verbeure, and P.Vets 
 as CLT below.
 In case of a single observable, their theorem is 
a kind of classical limit theorem for a family of spectral measures of operators appearing in quantum statistical mechanics.
However if we consider more than two observables,
the limit theorem suggests
the emergence of Boson fields as the algebra of fluctuation
of physical observables.
Then, it is natural to consider the time evolution
states of such Bose fields appearing in the central limit.
In fact, as the states appearing in CLT are quasifree states of Bose fields, at a heuristic level,
 the dynamics of the fluctuation operators is the quasifree 
dynamics (via Bogoliubov automorphisms) i.e. unitary time evolution in the test function space of the Boson field.
D.Goderis,  A.Verbeure,and P.Vets in \cite{Goderis3} investigated
the construction of the time evolution of the algebra of fluctuation.
They tried to introduce the time evolution for the algebra of fluctuation
induced by the microscopic dynamics .
As they could not prove their limit theorem to
any time invariant dense subalgebra, 
their introduction of the time evolution in an indirect way.
They have shown that the quasifree state of the algebra of fluctuation is a KMS state if the state of the microscopic system is KMS.
However they failed to prove validity of their assumption for
any non-trivial KMS states.
\par
After publication of \cite{Goderis1}, \cite{Goderis2}
and \cite{Goderis3}, no concrete example of quantum spin systems 
has been analyzed. 
 The difficulty lies in the mixing assumption in \cite{Goderis2}
where estimate of decay of correlation of two observables
$Q_1$ and $ Q_2$ located in two disjoint regions,
$\Lambda_1$ and  $\Lambda_2$.
In many cases we may prove the following estimate:
\begin{equation}
\abs{ \varphi (Q_1 Q_2 ) -  \varphi(Q_1)  \varphi(Q_2) }
\leq C(  Q_1, Q_2 ) \norm{Q_1}\norm{Q_2} r^{- d - \epsilon}
 \label{eqn:x1}
\end{equation}
where $r$ is the distance between the supports $\Lambda_1$
and $\Lambda_2$.
To show the central limit theorem,  we need information
of dependence of the constant $C(  Q_1, Q_2 ) $ on 
the size of  $\Lambda_1$ and  $\Lambda_2$.
D.Goderis, A.Verbeure, and P.Vets never proved their assumption
for any (non-product ) Gibbs state and until recently
it has been an open question whether the above results really hold
for any Gibbs states.
 \par
In \cite{XYCLT}, we obtained a  different proof of CLT of a single
 observable in quantum spin systems 
under a slightly different mixing condition . 
The advantages of our results in  \cite{XYCLT}
are as follows.
\newline
(i) We can prove our assumptions for several mixing states of one-dimensional systems.
The assumption of  \cite{XYCLT} is valid for 
 Gibbs states for finite range interactions , ground states
of the (massive) XY model, quasifree states of CAR algebras.
\newline
(ii) Our limit theorem is valid for (not strictly local)
certain quasi-local observables, which we named
{\em exponentially localized observables}.
\newline
In fact, instead of assuming (\ref{eqn:x1}), we can show CLT
under the following condition:
$$
\abs{ \varphi (Q_1 Q_2 ) -  \varphi(Q_1)  \varphi(Q_2) }
\leq C(  Q_1) \norm{Q_2} r^{- d -1- \epsilon}
$$
Our proof is based on an idea of E.Bolthauzen in \cite{Bolthauzen}. 
As far as strictly local observables concerned, 
we do not require any estimate of the constant $C( Q_1)$ and
we did not assume the assumption on the 4th moment 
(CLT4) of \cite{Goderis2} neither.
To show CLT for quasi-local observables
we assume stronger mixing , which is still valid
in many one-dimensional examples. 
\par
In this article,  we continue our study of CLT.
The new result of this article is 
 CLT for mutually non-commuting operators in the following sense:
$$ \lim_N \varphi (e^{iTQ^1_{<N>}}e^{iTQ^2_{N>}}....e^{iTQ^m_{<N>}})
=  e^{-\frac{t^2}{2} t(\sum_k Q^k , \sum_k Q^k )}
e^{ -i\frac{T^2}{2} \sum_{k<l} s( Q^k ,  Q^l ) }$$
This CLT is a stronger statement 
confirming appearance of Boson fields as the algebra of normal 
fluctuation.
We prove this limit theorem under the same condition of 
\cite{XYCLT} .  
As we are handling a convergence of functionals rather
that of measures, we had to add more argument and estimates
which are absent both in our previous paper \cite{XYCLT}
and in commutative cases.
Let us mention that Goderis, Verbeure and Vets have already
considered the same limit theorem
only for strictly local observables 
and again they could not prove their assumption
for any non-trivial Gibbs states. 
We will see that our CLT is valid for a dense algebra invariant
under the time evolution generated by
any finite range translationally invariant Hamiltonian.
As a corollary , we can introduce the time evolution for
the algebra of normal fluctuation directly.
The CLT of this paper implies that quasifree states
obtained in CLT is a KMS state of the algebra of fluctuation 
provided that the microscopic state satisfies the KMS condition
as well. 
\bigskip
\noindent
\par
We concentrate here one-dimensional system.
We may derive the same CLT for higher dimensional systems
under a similar mixing assumption of states, though the condition
looks rather difficult to verify , so we do not present its proof here.
In our proof of CLT for Gibbs states,  one dimensionality enters in 
use of the Ruelle transfer operator technique and 
the entire analyticity of the time evolution
(in Heisenberg Picture of Quantum Mechanics).
\bigskip
\noindent
\bigskip
\noindent
\par
Next we present our results more precisely.
We start with explaining our notation. 
Here the language of Operator Algebra is used and
the reader unfamiliar with this may consult 
\cite{BratteliRobinsonI} and \cite{BratteliRobinsonII}.
\par
By ${\calA}$ we denote the UHF
$C^*-$algebra $d^{\infty}$ (the infinite tensor product of d by d matrix
 algebras ) :
$${\calA} = \overline{\bigotimes_{\bfZ} \: M_{d}({\bf C})}^{C^*} .$$
Each component of the tensor product above is specified with a lattice
site $j \in \bfZ$.
By $Q^{(j)}$ we denote the element of ${\calA}_{\bfZ}$ with $Q$ in
the jth component of the tensor product and the unit in any other
component.
For a subset $\Lambda$ of $\bfZ$ , ${\calA}_{\Lambda}$ is defined as
the $C^*-$subalgebra of ${\calA}_{\bfZ}$ generated by elements 
localized in $\Lambda$.
When $\varphi$ is a state of ${\calA}_{\bfZ}$ the restriction
to ${\calA}_{\Lambda}$ will be denoted by ${\varphi}_{\Lambda}$ :
\begin{equation}
{\varphi}_{\Lambda} = \varphi \vert_{{\calA}_{\Lambda}} .
\label{eqn:a1}
\end{equation}
For simplicity we set 
\begin{equation}
\calA_{loc} = \cup_{\vert \Lambda\vert < \infty}
 {\calA}_{\Lambda}
\label{eqn:a2}
\end{equation}
where $\abs{\Lambda}$ is the cardinality of $\Lambda$.
Let $\tau_j$ be the lattice translation ( j shift to the right) determined 
by 
\begin{equation}
\tau_j(Q^{(k)})=Q^{(j+k)}
\label{eqn:a3}
\end{equation}
 for any j and k in $\bfZ$.
\bigskip
\noindent
\bigskip
\noindent
\newline
{\bf@Time Evolution of Local Observables}
\newline
In this paper, the time evolution of (microscopic) local observables is 
determined by a translationally invariant finite range interaction. 
Our local Hamiltonian $\hal$ on a finite volume $\Lambda$ has the following standard form:
\begin{equation}
\hal =\sum_{X \subset \Lambda} \psi (X)
\label{eqn:a4}
\end{equation}
where $X$ is an interval $[a,b]$ of $\bfZ$ and
$$\psi (X) =\psi (X)^* \in \calA_X  \quad , \quad 
\tau_j (\psi (X)) = \psi (X+j) $$
\begin{equation}
\psi (X) =0   \quad \quad \mbox{if the diameter $d(X)$ of the set $X$ is greater than $r$.}
\label{eqn:a5}
\end{equation}
The  time evolution $\evl^{\Lambda}$
of the finite system in a finite set $\Lambda$ of $\bfZ$  
is determined by
$$\evl^{\Lambda}(Q) = e^{it\hal}Q e^{-it\hal} .$$
From the local time evolution $\evl^{\Lambda}$,
 the global time evolution $\evl$ is obtained via the thermodynamic
limit which converges in the norm topology of $\calA$ .
$$ \evl (Q) = \lim_{\Lambda \to \bfZ} \evl^{\Lambda}(Q) .$$
Next,we introduce the notion of {\em exponentially localized observables} .
\begin{df} 
Let $\theta$ be a positive constant $0< \theta<1$ .
Define $\norm{Q }^{(n)}$  by the following equation:
\begin{equation}
\norm{Q }^{(n)} = 
\inf \cbk{ \norm{Q - Q_n}  \; \vert \; Q \in \calA_{[-n,n]} } 
\label{eqn:a6}
\end{equation}
for n positive, $n > 0$ and we set
$$\norm{Q }^{(0)}=\norm{Q} .$$
In terms of $\norm{Q }^{(n)}$  we introduce $\tri{Q}_{\theta}$ :
\begin{equation}
\tri{Q}_{\theta} = \sum_{n=0,1,2, ... } \norm{Q }^{(n)} \theta^{-n} 
\label{eqn:a7}
\end{equation}
 An element $Q$ of $\calA$ is exponentially localized with rate $\theta$
if $\tri{Q}_{\theta}$ is finite.
\par
The set of all exponentially localized elements with rate $\theta$
is denoted by $F_{\theta}$.
\par
We fix an element $Q_n$ of  $\calA_{[-n,n]}$  satisfying 
 $$\norm{Q }_{n} =  \norm{Q - Q_n } . $$
\end{df} 
Note that due to compactness of a closed bounded subset of  
$\calA_{[-n,n]}$ the infimum in (\ref{eqn:a6}) is attained by an element . 
In \cite{TM2} we used slightly different definition of $F_{\theta}$.
But they are essentially equivalent.  
\par
The following invariance of $F_{\theta}$ under the time evolution
is a corollary of results due to H.Araki
in \cite{Araki} or due to the {\em Propagation of Estimates}
due to E.Lieb and D.Robinson
(c.f. Theorem 6.2.9. and 6.2.11 of \cite{BratteliRobinsonII}) .
\begin{pro}
Suppose that the interaction is of finite range.
\newline
(i) For any $\theta$ and $\theta'$ with $\theta < \theta'$,
there exists a constant $C = C(\theta ,\theta')$
 such that
$$ \tri{\evl (Q) }_{\theta'} \leq C \tri{Q}_{\theta} $$
(ii) For any $Q$ in $F_{\theta}$,
 $\evl(Q)$ is entire analytic (analytic in the whole complex plane) 
as a function of $t$. 
\end{pro}
Due to the above proposition the sets
$\cup_{\theta} F_{\theta}$ and 
$\cap_{\theta} F_{\theta}$ are invariant under $\evl$.
\bigskip
\noindent
\bigskip
\noindent
\newline
{\bf Gibbs states}
\newline
As usual, the local Gibbs state at the inverse temperature $\beta$ 
is defined by
\begin{equation}
\varphi_{\beta}^{(\Lambda)}(Q) = 
\frac{tr(e^{-\beta \hal} Q)}{tr(e^{-\beta \hal})} .
\label{eqn:x2}
\end{equation}
For the finite range interaction, the following thermodynamics limit 
exists: 
$$ \lim_{N \to \infty} \varphi_{\beta}^{([-N,N])}(Q)
=\varphi_{\beta}(Q).$$
$\varphi_{\beta}$ is  a unique Gibbs state of the time evolution
 $\evl$ and $\varphi_{\beta}$ satisfies the KMS boundary condition:  
$$\varphi_{\beta}(Q_1 \alpha_{i\beta}(Q_2)) = 
\varphi_{\beta}(Q_2 Q_1)$$ 
for $Q_1$ and  $Q_2$ in $F_{\theta}$. 
By applying the Ruelle transfer operator technique for UHF algebras
(c.f. \cite{Araki}, \cite{Golodets} , and \cite{TM2})
we obtain the following estimates for the infinite volume Gibbs state
$\varphi_{\beta}$.
\begin{theorem}
Suppose that the interaction is of finite range.
\newline
(i) There exist constants $C, m>0$ such that
\begin{equation}
\abs{\varphi_{\beta}^{[-k+a,b+k]}(Q)-\varphi_{\beta}(Q)}
\leq C e^{-m k}\norm{Q}
\label{eqn:a11}
\end{equation}
for any  $Q$ in $A_{[a,b]}$.
\newline
(ii) We have exponential decay of correlation for observables
localized in the half infinite intervals $(-\infty ,-1]$ and, $[0,\infty)$.
\begin{equation}
\abs{\varphi(Q_1 \tau_j(Q_2)) -
\varphi(Q_1 )\varphi(Q_2)} \leq K e^{-M j }
\norm{Q_1}\norm{Q_2}
\label{eqn:a12}
\end{equation}
for   $Q_1$ in ${\calA}_{(-\infty, -1]}$, $Q_2$ in ${\calA}_{[0,\infty)}$ 
and $j >0$.
\end{theorem}
\bigskip
\noindent
\bigskip
\noindent
\newline
{\bf Central Limit Theorem}
\newline
Suppose that $\varphi$ is a translationally invariant factor state.
Consider the local fluctuation $Q_{<N>}$ of an observable $Q$ :
$$Q_{<N>}= \frac{1}{\sqrt{2N+1}} \rbk{\sum_{\abs{j}\leq N}
(\tau_j(Q)- \varphi(Q)) } .$$
If the observable is diagonal, the limit $\lim_N Q_{<N>}$ makes
 sense as convergence of spectral measures and this is nothing 
but a classical central limit theorem of mixing systems. 
The central limit does not converge in neither strong 
nor weak topology of operators in a Hilbert space. 
Nevertheless,
the following weak limit exists in the GNS space of 
$\varphi$.
\begin{equation}
w-\lim_{N\to \infty} \sbk{ Q_{<N>} , R_{<N>} } = s(Q,R) 1
\label{eqn:a13}
\end{equation}
where
\begin{equation}
s(Q,R) = \sum_{k \in \bfZ} \varphi ( \sbk{\tau_k(Q) ,R} ) .
\label{eqn:x4}
\end{equation}
Set
\begin{equation}
t(Q,R) = \lim_{N \to \infty} \varphi (Q_{<N>} R_{<N>}) . 
\label{eqn:a14}
\end{equation}
\begin{df}
Let $\varphi$ be a translationally invariant state of $\calA$.
Suppose that  $Q$ in $\calA$ is selfadjoint.
We say that the central limit theorem holds for $Q$ and 
$\varphi$ if 
\begin{equation}
\lim_{N \to \infty} \varphi(e^{iT Q_{<N>}}) =
e^{-\frac{T^2}{2} t(Q,Q)}
\label{eqn:a15}
\end{equation}
\end{df}
\begin{theorem}
(i) Let $\varphi$ be a translationally invariant state .
Suppose there exist  positive constants $M(j)$  such that
$$\sum_{j=1}^{\infty} j M(j)  < \infty$$
and
$$\abs{\varphi(Q_1 \tau_j(Q_2)) -
\varphi(Q_1 )\varphi(Q_2)} \leq K \norm{Q_1}\norm{Q_2} M(j)$$
for  $Q_1$ in ${\calA}_{(-\infty, -1]}$, $Q_2$ in ${\calA}_{[0,\infty)}$ 
and $j >0$.
\par
The central limit theorem holds for any  selfadjoint strictly
local observable $Q$ in $\calA_{loc}$. 
We have convergence of the following correlation functions
for local $Q(k)$ ($k=1,2,...r$).
\begin{equation}
\lim_{N \to \infty} \varphi(\prod_{k=1}^{r} e^{i Q(k)_{<N>}} ) =
e^{-\frac{1}{2} t(\sum_{k} Q(k),\sum_{k} Q(k))}
e^{ - \frac{i}{2} \sum_{k<l} s(Q(k),Q(l))}
\label{eqn:a16}
\end{equation}
(ii) If the translationally invariant state satisfies the exponential mixing 
condition (\ref{eqn:a12}) , (\ref{eqn:a16}) is valid for
for any exponentially localized selfadjoint $Q(k)$ in  $F_{\theta}$. 
\end{theorem}
The convergence of (\ref{eqn:a16}) was not proved in \cite{XYCLT} .
Our proof is similar to the case of a single observable. 
We will give our proof in Section 3.
\bigskip
\noindent
\bigskip
\noindent
\newline
{\bf Time Evolution of the Algebra of Fluctuation}
\newline
As the Gibbs state $\varphi_{\beta}$ for  any finite range interaction
 satisfies the assumption of Theorem 1.5, 
we obtain a direct construction of dynamics
for the algebra of fluctuation.
The set of all selfadjoint exponentially localized elements
with rate $\theta$ is 
denoted by $F_{\theta}^{real}$.
Set
$$ \calF = \cup_{0< \theta <1 } \:\:@F_{\theta}^{real} .$$
Due to our assumption on decay of correlation, 
\begin{equation}
t(Q,Q)= \sum_{k \in \bfZ} \cbk{ \varphi_{\beta}(Q\tau_k(Q)) - 
\varphi_{\beta}(Q)^2 } 
\label{eqn:a17}
\end{equation}
is finite and non-negative.
Then, $t(Q,Q)$ of (\ref{eqn:a14}) is a (degenerate) positive semidefinite 
bilinear form. We denote the kernel of $t$ by $\calN$.
$$\calN = \cbk{ Q \in \calF  \:\: \vert \:\:\: t(Q,Q) =0 }$$
Next we introduce the equivalence relation on 
$\calF$ . $A$ and $B$ are equivalent if $A-B$ is in $\calN$.
By $\tilde{A}$ we denote the equivalence class for $A$ in 
$\calF$. Then $s(A,B)$ of (\ref{eqn:a13})
gives rise to a non degenerate symplectic form 
$\tilde{s}(\tilde{A},\tilde{B})$ on $\tilde{F}$
where
\begin{equation}
\tilde{F} = \calF / \calN 
\label{eqn:a18}
\end{equation}
and
\begin{equation}
\tilde{s}(\tilde{A},\tilde{B}) = s(A,B) .
\label{eqn:a19}
\end{equation}
As was already discussed in \cite{Goderis2} and \cite{Goderis3}, 
the central limit gives rise to the Weyl algebra $W(\tilde{F} , \tilde{s})$
on the symplectic space $\cbk{  \tilde{F} , \tilde{s} }$. 
\par
More precisely, we define ${\cal W}(\tilde{F} , \tilde{s})$ by the $C^*$-algebra generated by unitaries $W(\tilde{Q})$ satisfying
\begin{equation}
W(\tilde{Q_1})W(\tilde{Q_2})=
W(\tilde{Q_1}+\tilde{Q_2})e^{-i1/2 s(\tilde{Q_1},\tilde{Q_2})}.
\label{eqn:a20}
\end{equation}
${\cal W}(\tilde{F} , \tilde{s})$ is called 
the {\em algebra of normal (macroscopic) fluctuations}.
\par
When the interaction is of finite range,
 the global dynamics of $\calA$ leaves  $\calF$ invariant
and we can introduce the dynamics $\tilde{\alpha}_t$ on 
the Weyl algebra ${\cal W}(\tilde{F} , \tilde{s})$
via the following equation:
\begin{equation}
\tilde{\alpha}_t (W(\tilde{Q}) = W(\tilde{\evl (Q)}) .
\label{eqn:a21}
\end{equation}
The time evolution $\tilde{\alpha}_t$ has weak continuity in the sense
that $\tilde{\evl (Q)}$ is continuous with respect to the topology
induce by the bilinear form $t(Q,Q)$.
In fact, if  we have exponential clustering (\ref{eqn:a12}) 
, it is not difficult to show
\begin{equation}
\lim_{u \to 0} t(\alpha_u (Q) -Q,\alpha_u (Q) - Q) =0 .
\label{eqn:a22}
\end{equation}
Note that the left-hand side of (\ref{eqn:a22}) is equal to
$$\lim_{t \to 0}
\sum_{k \in \bfZ} \varphi_{\beta}(\{ \evl (Q) -Q , \tau_k (Q) \} ) =0 .$$
As was argued in \cite{Goderis2},
the central limit $\lim_N \varphi (e^{iQ_{<N>}})$ defines
a quasifree state $\tilde{\varphi}$ of the algebra of normal 
fluctuations ${\cal W}(\tilde{F} , \tilde{s})$.
\par
We now consider the GNS representation of the algebra of normal 
fluctuations ${\cal W}(\tilde{F} , \tilde{s})$ associated with
$\tilde{\varphi}_{\beta}$ .
Let $\{ \pi ({\cal W}(\tilde{F} , \tilde{s})), \Omega , \calH \}$ be the
GNS triple for  $\tilde{\varphi}_{\beta}$ .
Let $\calM$ be the von Neumann algebra generated by
$\pi ({\cal W}(\tilde{F} , \tilde{s}))$.
As  $\tilde{\varphi}_{\beta}$ is invariant under the time evolution
$\tilde{\alpha}_t$ of the algebra of normal 
fluctuations ${\cal W}(\tilde{F} , \tilde{s})$ , we have a one parameter  
group of unitary $U_t$ on $\calH$ which implements $\tilde{\alpha}_t$.
$$U_t \pi (Q) U_t^* = \pi (\tilde{\alpha}_t(Q)) \quad , \quad 
U_t \Omega =\Omega .$$
As a consequence of the continuity (\ref{eqn:a22}), we see that
the adjoint action of $U_t$ gives rise to a weakly continuos
one parameter group of automorphisms of $\calM$.
We denote this dynamics of $\calM$ by the same symbol
$\tilde{\alpha}_t$:
$$ \tilde{\alpha}_t(M) = U_t M U_t^*$$
for $M$ in $\calM$.
We state our main result of this paper.
\begin{theorem}
Let ${\varphi}_{\beta}$ be the unique $\beta$-KMS state for
a finite range translationally invariant interaction of a one-dimensional
quantum spin chain. The quasifree state
$\tilde{\varphi}_{\beta}$ is 
a $\beta$-KMS state for the dynamics
$\tilde{\alpha}_t$ of the von Neumann algebra $\calM$ at the same inverse temperature $\beta$.
\end{theorem}
It is easy to derive the KMS boundary condition of
$\tilde{\varphi}_{\beta}$ from 
the central limit theorem (\ref{eqn:a16} for mutually non-commuting
observables.
Note that the argument of \cite{Goderis3} is based on a weaker version
CLT i.e. CLT only for strict local observables, which makes arguments
more complicated.
\bigskip
\noindent
\bigskip
\noindent
\newline
We explain the contents of the rest of this paper now.
Section 2 is a mathematical preliminary.
In Section 3, we present our proof of Theorem 1.5
(\ref{eqn:a16}) and that of Theorem 1.6.
\par
In Section 4 we make a few comment on unsolved problems.
Here, we present our proof of the uniform exponential 
mixing condition for finitely correlated states.
Even  though the result is as expected,
we are not aware of  any proof  published elsewhere.
Our proof is an simple application of {\em dual}
transfer operator .

\section{Localization }
\setcounter{theorem}{0}
\setcounter{equation}{0}
We present some features of
exponentially localized elements.
\begin{lmm}
$F_{\theta}$ is complete in the norm topology induced by
 $\tri{Q}_{\theta}$ . 
Any  bounded set of $F_{\theta}$  is a compact subset of $\calA$ in norm topology.
\end{lmm}
\begin{lmm}
 Let $Q$ be exponentially localized with rate $\theta$.
There exists an element $\tilde{Q}_n$ 
of  $\calA_{[-n,n]}$  satisfying
\begin{equation}
\norm{Q - \tilde{Q}_n } \leq 2\tri{Q}_{\theta} \theta^n
\; , \; 
\norm{\tilde{Q}_n} \leq \norm{Q}
\label{eqn:a8}
\end{equation}
If $Q$  is positive we can find a positive
$\tilde{Q}_n$ such that
\begin{equation}
0 \leq \inf Q \leq \inf \tilde{Q}_n \leq \norm{\tilde{Q}_n} \leq \norm{Q}
\label{eqn:a9}
\end{equation}
where $\inf Q$ is the infimum of the spectrum of $Q$.  
\end{lmm}
One choice of  $\tilde{Q}_n$ is
${tr}_{(\infty , -n]\cup [n, \infty)}(Q)$ where
${tr}_{(\infty , -n]\cup [n, \infty)}$ is the partial trace 
(conditional expectation to $\calA_{[-n,n]}$).
\begin{lmm}
Let $Q^1$ and $Q^2$ be exponentially localized,
$Q^1 , Q^2 \in F_{\theta}$. Then the following sum is finite.
\begin{equation}
\sum_{k \in \bfZ} \norm{\sbk{\tau_{k}(Q^1) , Q^2}} < \infty
\label{eqn:b1}
\end{equation}
\end{lmm}
{\it Proof}
\newline
Fix  $Q^i_n$ in $\calA_{[-n,n]}$
satisfying
$$\norm{Q^i - Q^i_n} =\norm{Q }^{(n)} .$$
for $i=1,2$ and set
$$R^i_n = Q^i_n - Q^i_{n-1}  \:\: ( n \geq 1)
\quad , \quad R^i_0 = Q^i_0 .$$
Then
$$ \norm{R^i_n} \leq \norm{Q^i_n} + \norm{ Q^i_{n-1}}
\leq  (1 + \theta^{-1} ) \theta^n \tri{Q^i}_{\theta} .$$
Now we have
$$Q^i = \sum_{n=0}^{\infty} R^i_n .$$
As a consequence,
\begin{eqnarray}
\sum_{k \in \bfZ} \norm{\sbk{\tau_{k}(Q_1) , Q_2}}
& &\leq \sum_{n,m =0}^{\infty}\sum_{k \in \bfZ} 
\norm{\sbk{\tau_{k}(R^1_n) , R^2_m}}
\nonumber\\
& &\leq 2 \sum_{n,m =0}^{\infty} (2(m+n)+1)\norm{R^1_n}\norm{R^2_m}
\nonumber\\
& &\leq 2 \sum_{n,m =0}^{\infty}  
(1 + \theta^{-1} )^2 \theta^n  \theta^m \tri{Q^1}_{\theta} 
\tri{Q^2}_{\theta} <\infty
\label{eqn:b2}
\end{eqnarray}
{\it End of Proof}
\bigskip
\noindent
\bigskip
\noindent
\par
The following result is due to H.Araki in \cite{Araki}.
\begin{pro}
Suppose that the interaction is of finite range.
For any $Q$ in $F_{\theta}$,
 $\evl(Q)$ is entire analytic
(analytic in the whole complex plane) as a function of $t$. 
\par
For any $\beta > 0$ and any $\theta$, 
there exists a constant $M=M(\beta , \theta , r )$ independent of 
the interaction $\psi$
such that the following estimate is valid:
\begin{equation}
\norm{\alpha_z(Q)}_{\theta}\leq  M 
F_0(2 \abs{\beta}\norm{h(\psi)})  
\tri{Q}_{\theta e^{4\abs{\beta}\norm{h(\psi)}}} 
\label{eqn:b3}
\end{equation}
where any complex number $z$ satisfying $\abs{z} \leq \beta$ and
$$h(\psi ) = \sum_{X : 0 \in X } \frac{\psi(X)}{d(X)} , $$
$$F_0(x)=exp[(-r+1) x + 2 \sum_{k=1}^r k^{-1}(e^{kr} -1)] .$$
\end{pro}
Thus this proposition tells us that
$\cap_{0<\theta <1} F_{\theta}$ is
a dense $\alpha_z$ invariant subalgebra of $\calA$.
\par
Next we show that
$\cup_{0<\theta <1} F_{\theta}$ is a dense $\alpha_t$ invariant
subalgebra of $\calA$.
\bigskip
\noindent
\newline
{\em Proof of Proposition 1.2 (i).}
The proof of Proposition 6.2.9 of \cite{BratteliRobinsonII}.
tells us the following estimate:
\begin{equation}
\norm{\evl (A) - e^{it \hal} A e^{-it \hal}} \leq
\norm{A} \abs{\Lambda_0} ( e^{2\abs{t} \norm{\psi}_{\lambda}} -1)
\sum_{j \in \Lambda^c } e^{-\lambda d(j, \Lambda_0)}
\label{eqn:b4}
\end{equation}
where $A$ is localized in $\Lambda_0$ , $\lambda$ is any positive 
constant,  $d(j, \Lambda_0)$ is the distance of $j$
from $\Lambda_0$ and $\norm{\psi}_{\lambda}$
is a positive constant depending on the interaction $\psi$ and 
$\lambda$ .
\par
Now take $Q$ from $F_{\theta}$ and 
in (\ref{eqn:b4}),we set $e^{-\lambda} = \theta_0$
where $ \theta < \theta_0 < \theta' $.
 $A=R_m = Q_m - Q_{m-1}$ in $\calA_{ [-m, m]}$ , $\Lambda_0 = [-n,n]$. 
For $n$ larger than $m$, we obtain
\begin{eqnarray*}
\norm{\evl (R_m)}^{(n)} &\leq& 
2( e^{2\abs{t} \norm{\psi}_{\lambda}} -1)(2m+1) \norm{R_m} 
\sum_{k=n}^{\infty}  \theta_0^{k - m}
\\ 
&\leq&
2 ( e^{2\abs{t} \norm{\psi}_{\lambda}} -1)(2m+1) 
\theta^{n - m} \frac{\norm{R_m}}{1- \theta_0}
\end{eqnarray*}
On the other hand , in general,
$$\norm{Q}^{(n)} \leq 2 \norm{Q}$$ 
due to Lemma 2.2.
As a consequence, for $n$ smaller than $m$, $ n < m$,
 we have
$$ \norm{\evl (R_m)}^{(n)} 
\leq 2 \norm{R_m} \leq 4 \theta^n \tri{Q}_{\theta} .$$ 
Then,
\begin{eqnarray}
& &\tri{\evl (Q) }_{\theta'} \leq
\sum_{m=1}^{\infty}   \tri{\evl (R_m) }_{\theta'} 
\nonumber\\
&\leq& \sum_{m=1}^{\infty} 
\cbk{ \sum_{k =0}^{m} {\theta'}^{-k} \norm{\evl (R_m)}^{(k)}
+ 2 ( e^{2\abs{t} \norm{\psi}_{\lambda}} -1)
\frac{ (2m+1)}{1-\theta_0}  
 \norm{R_m} \sum_{k=m+1}^{\infty} {\theta'}^{-k}\theta_0^{k - m} }
\nonumber\\
&\leq&
C_1 \sum_{m=1}^{\infty} 
4(m+1)( \frac{\theta'}{\theta_0})^{-m} \theta_0^{-m} \norm{R_m} 
\nonumber\\
&\leq& C_2
\sum_{m=1}^{\infty} {\theta_0}^{-m} \norm{R_m}
=C_3 \tri{Q}_{\theta}
\label{eqn:b5}.
\end{eqnarray}
Note that
$ \sup_m ( (2m+1) {\theta'}^{-m} {\theta_0}^{m} )$ is finite.
{\em End of Proof}

\section{Proof of Theorem 1.5 and 1.6.}
\setcounter{theorem}{0}
\setcounter{equation}{0}
We present here our proof for the convergence of two point 
correlation for (\ref{eqn:a16}) .
We concentrate the case of exponentially localized observables
as this is what we need to define the time evolution and KMS states
for the algebra of normal fluctuation.
\par
The idea of proof is as follows. Set
\begin{equation}
 F^{N}_{(Q,R)}(T) =
\varphi (e^{i TQ_{<N>}}e^{i T R_{<N>}}) . 
\label{eqn:d1}
\end{equation}
Instead of showing
$$\lim_{N \to \infty} F_N(Q,R)=
e^{-1/2 T^2 t(Q+R,Q+R) }  e^{ -i/2 T^2 s(Q,R)}) , $$
we prove that the following limit vanishes. 
\begin{equation}
\lim_{N \to \infty} \rbk{ \frac{d}{dT} F_N(Q,R)(T) 
+ T( t(Q+R,Q+R) + i s(Q,R)) F_N(Q,R)(T) }=0 .
\label{eqn:d2}
\end{equation}
Note that the above claim is same as 
showing the following :
\begin{eqnarray*}
\lim_{N \to \infty} & & 
\{ \varphi (( T(it(Q+R,Q+R)- s(Q,R)) - Q_{<N>}) 
e^{i TQ_{<N>}}e^{i T R_{<N>}} ) 
\\
& &- \varphi (e^{i TQ_{<N>}} R_{<N>}e^{i T R_{<N>}}) \}=0 .
\end{eqnarray*}
If the second derivative of $F^{N}_{(Q,R)}(T)$ is  bounded in a small
neighborhood of the origin uniformly in N ,  $F^{N}_{(Q,R)}(T)$ and its derivative are equi-continuous. It turns out that
 there is at least one accumulation point  $F^{\infty}_{(Q,R)}(T)$
in the sequence of the functions $F_N(Q,R)(T)$ ($ N=1,2,3,...$).
Due to equicontinuity of the first and the second derivative of 
$F^{N}_{(Q,R)}(T)$
, $F^{\infty}_{(Q,R)}(T)$  is differentiable.
Then (\ref{eqn:d2}) implies $F^{\infty}_{(Q,R)}(T)$ satisfies
\begin{equation}
\frac{d}{dT}F^{\infty}_{(Q,R)}(T) = T ( t(Q+R,Q+R)-is(Q,R) )
F^{\infty}_{(Q,R)}(T) .
\label{eqn:d3}
\end{equation}
The unique solution to (\ref{eqn:d3}) is 
$e^{-1/2 T^2 t(Q+R,Q+R) }  e^{ -i/2 T^2 s(Q,R)}$ .
\bigskip
\noindent
\par
Now we begin the proof of the above claim.
Without loss of generality
we assume that
$$\varphi(Q)=\varphi(R)=0$$
 \par
Suppose that $A=A^*$ $B=B^*$ are selfadjoint bounded operators
on a Hilbert space. Then
\begin{eqnarray}
& &\norm{e^{i(A+B)} -e^{iB} e^{iA}} \leq \norm{ \sbk{ A , e^{iB}} }
\nonumber\\
& &\norm{ \sbk{ A , e^{iB} }} \leq \norm{\sbk{ A ,B}} 
\label{eqn:d4}
\end{eqnarray}
(c.f.\cite{Goderis2} or \cite{XYCLT}.)
\begin{lmm} 
The first and the second derivative of $F^{N}_{(Q,R)}(T)$ are  bounded
uniformly in $N$.
\end{lmm}
{\it Proof :}
The second derivative $F^{N}_{(Q,R)}(T)$ has the following expression:
\begin{eqnarray}
-\frac{d^2}{dT^2} F^{N}_{(Q,R)}(T)
\nonumber\\
 =& &
\varphi ({Q_{<N>}}^2 e^{i TQ_{<N>}}e^{i T R_{<N>}} ) +
\varphi (e^{i TQ_{<N>}}e^{i T R_{<N>}}{R_{<N>}}^2) 
\nonumber\\
 & &+
2\varphi(Q_{<N>}e^{iT Q_{<N>}}e^{i T R_{<N>}} R_{<N>}) .
\label{eqn:d5}
\end{eqnarray}
Each term in (\ref{eqn:d5}) is uniformly bounded in $N$.
We focus on $\varphi ({Q_{<N>}}^2 e^{i TQ_{<N>}}e^{i T R_{<N>}} )$.
\par
Due to exponential decay of correlation, we have
$$\lim_{N \to \infty} \varphi ({Q_{<N>}}^2) = \sum_{k\in \bfZ}
\varphi(Q \tau_k(Q))  < \infty $$
where we used $\varphi(Q)=0$.
Thus we have a constant $K_1$ such that for any $N$
$$0 \leq \varphi ({Q_{<N>}}^2)  \leq K_1 .$$
By use of (\ref{eqn:d4}) we can show  uniform boundedness
of the commutator of $Q_{<N>}$ and $e^{i T R_{<N>}}$ :
\begin{eqnarray}
\norm{ \sbk{ Q_{<N>} , e^{i T R_{<N>}} }}
& &\leq T^2 \norm{ \sbk{ Q_{<N>} ,  R_{<N>}} }
\leq  T^2 \frac{1}{2N+1} \sum_{0 \leq \abs{a} ,\abs{b} \leq N}
\norm{ \sbk{ Q , \tau_{b-a}(R) } }
\nonumber\\
& &\leq  T^2 \sum_{k \in \bfZ }
\norm{\sbk{ Q , \tau_{k}(R) } } \leq K_2 < \infty .
\label{eqn:d6}
\end{eqnarray}
Then,
\begin{eqnarray*}
& &\abs{\varphi ({Q_{<N>}}^2 e^{i TQ_{<N>}}e^{i T R_{<N>}} )}
\\
\leq & &
\abs{\varphi (Q_{<N>}e^{i TQ_{<N>}}e^{i T R_{<N>}} Q_{<N>}) }
+ \abs{ \varphi (Q_{<N>}e^{i TQ_{<N>}}\sbk{ e^{i T R_{<N>}}, Q_{<N>} }) }
\\
\leq & &
\varphi (Q_{<N>}^2 )^{1/2}
\rbk{\varphi ( (e^{i T R_{<N>}} Q_{<N>})^* (e^{i T R_{<N>}} Q_{<N>}))^{1/2} 
+ \norm{ \sbk{ Q_{<N>} , e^{i T R_{<N>}} }} }
\\
\leq & & K_1 + ( K_1)^{1/2} K_2  < \infty
\end{eqnarray*}
By the same estimate,we see that other terms of (\ref{eqn:d5})
and the first derivative of $F^{N}_{(Q,R)}(T)$ are uniformly bounded in $N$. 
{\it End of Proof}
\begin{lmm} 
\begin{equation}
\lim_{n \to \infty} 
\varphi 
( (i Ts(Q,R) + ( e^{i TQ_{<N>}} R_{<N>}e^{-i TQ_{<N>}}-   R_{<N>} ) )
e^{i TQ_{<N>}} e^{i T R_{<N>}})=0
\label{eqn:d7}
\end{equation}
\end{lmm}
{\it Proof:}
First note that
\begin{eqnarray}
& &\norm{ e^{i TQ_{<N>}} R_{<N>}e^{-i TQ_{<N>}}-   R_{<N>} -
iT\sbk{ Q_{<N>} , R_{<N>} } }
\nonumber\\ 
&\leq & C T^2  \norm{ \sbk{Q_{<N>} , \sbk{ Q_{<N>} , R_{<N>}} }} .
\label{eqn:d8}
\end{eqnarray}
The right-hand side of (\ref{eqn:d8}) is bounded from above by
\begin{eqnarray}
& &C T^2 \frac{1}{(2N+1)^{3/2} } \sum_{ \abs{i} , \abs{j} ,\abs{k} \leq N}
\norm{ \sbk{ \tau_{j-k} (Q) , \sbk{  \tau_{i-k} (Q) , R } }} 
\nonumber\\
\leq & &C T^2 \frac{1}{(2N+1)^{1/2} } \sum_{ \abs{i} , \abs{j}  \leq 2N}
\norm{ \sbk{ \tau_{j} (Q) , \sbk{  \tau_{i} (Q) , R } }}
\label{eqn:d9}
\end{eqnarray}
If $Q$ and $R$ are local, the non vanishing terms in the right-hand side 
of (\ref{eqn:d9}) are finite: 
$$\sbk{ \tau_{j} (Q) , \sbk{  \tau_{i} (Q) , R } }=0$$
for $\abs{i} , \abs{j} \geq r$.
When $Q$ and $R$ are exponentially localized, the commutator
 $\sbk{  \tau_{i} (Q) , R }$ is also exponentially localized and
$$\norm{ \sbk{ \tau_{j} (Q) , \sbk{  \tau_{i} (Q) , R } }} \leq C
\theta^{\abs{j}+\abs{i}} .$$
Thus
$$\sum_{i, j \in \bfZ }
\norm{ \sbk{ \tau_{j} (Q) , \sbk{  \tau_{i} (Q) , R } }} < \infty .$$
and we obtain the following convergence:
\begin{equation}
\lim_{N \to \infty}
\norm{ e^{i TQ_{<N>}} R_{<N>}e^{-i TQ_{<N>}}-   R_{<N>} -
iT\sbk{ Q_{<N>} , R_{<N>} } } =0 .
\label{eqn:d10}
\end{equation}
Finally we show the following convergence which implies
the claim of Lemma 3.2.
\begin{equation}
\lim_{N \to \infty} 
\varphi ( ( s(Q,R) - \sbk{ Q_{<N>} , R_{<N>} })
e^{i TQ_{<N>}} e^{i T R_{<N>}})=0
\label{eqn:d11}
\end{equation}
Then, 
$$\sbk{ Q_{<N>} , R_{<N>} }=\frac{1}{2N+1}\sum_{\abs{i}\leq N}
\sum_{\abs{j}\leq N} \tau_i (\sbk{ \tau_{j-i }(Q) , R})$$
and
\begin{equation}
\lim_{N \to \infty} \sum_{\abs{j}\leq N} 
\sbk{ \tau_{j-i }(Q) , R} = \sum_{j \in \bfZ} 
\sbk{ \tau_{j}(Q) , R}.
\label{eqn:d12}
\end{equation}
Note that the right-hand side of (\ref{eqn:d12}) converges in the norm
topology due to exponential localization.
We define $A$ via the following equation:
$$A = \sum_{j \in \bfZ} \sbk{ \tau_{j}(Q) , R}.$$
Note that
$$s(Q,R) = \varphi (A) .$$ 
Instead of showing (\ref{eqn:d11}) , we have only to prove
\begin{equation}
\lim_{N \to \infty} 
\varphi ( \rbk{ s(Q,R) - \lim_{N \to \infty} \frac{1}{2N+1}
\{ \sum_{\abs{j}\leq N} \tau_j (A) \} } e^{i TQ_{<N>}} e^{i T R_{<N>}})=0
\label{eqn:d13}
\end{equation}
It is not difficult to see that
the following identity implies (\ref{eqn:d13}).
\begin{equation}
\lim_{N \to \infty}  \varphi ( ( \frac{1}{2N+1}
\sum_{\abs{j}\leq N} \tau_j (A)   - \varphi (A))^2 ) =0.
\label{eqn:d14}
\end{equation}
 The left-hand side of (\ref{eqn:d14}) is equal to 
\begin{eqnarray}
\lim_{N \to \infty}  \frac{1}{(2N+1)^2}
& &\sum_{\abs{i} , \abs{j} \leq N} 
\varphi ( \tau_{i-j} (A) A ) - \varphi (A)^2
\nonumber\\
=& &\lim_{N \to \infty}  \frac{1}{(2N+1) }
\sum_{\abs{i} \leq N} 
\varphi ( \tau_{i} (A) A) - \varphi (A)^2  =0.
\label{eqn:d15}
\end{eqnarray}
The last identity is due to ergodicity of the state $\varphi$.
(\ref{eqn:d15}) implies the claim of Lemma.
{\it End of Proof}
\bigskip
\noindent
\bigskip
\noindent
\par
We can find an increasing sequence $m(N)$ of positive integers such that 
\begin{equation}
\lim_{N \to \infty} e^{-m(N)} \sqrt{2N+1} =0 \: ,\:
\lim_{N \to \infty} m(N)^{-1} (2N+1)^{1/4} =\infty .
\label{eqn:d16}
\end{equation}
We set
\begin{eqnarray*}
Q_{j,N}= & &\sum_{ \abs{k}\leq N , \abs{k-j}\leq  m(N)} \tau_k(Q)
\quad , \quad
\beta_N = \sum_{ \abs{k}\leq N } 1/2\varphi(\{\tau_k (Q) \: , \: Q_{k,N}\})
\\
 \overline{Q}_{<N>} & &=
\beta^{-1/2}_N \cbk{\sum_{ \abs{k}\leq N } \tau_k(Q)}  \quad , \quad
\overline{Q}_{j,N} = \beta^{-1/2}_N  Q_{j,N} .
\end{eqnarray*}
Then $\beta_N=(2N+1)(1+o(1))$ due to summability of two point  correlation functions. Thus we have only to consider
 $$\varphi (e^{i T\overline{Q}_{<N>}}e^{i T \overline{R}_{<N>}})$$
and its derivative.
\par
Furthermore, in what follows, we assume that $t(Q+R,Q+R) =1$
for simplicity of exposition. ( The case $t(Q+R,Q+R) =0$ can be handled
in the same manner.)
We set
$$ P = Q+R.$$
\begin{lmm} 
\begin{equation}
\lim_{n \to \infty} \varphi((iT-\overline{P}_{<N>})
e^{iT\overline{Q}_{<N>}}e^{iT\overline{R}_{<N>}})
=0
\label{eqn:d17}
\end{equation}
\end{lmm}
{\it Proof:}
We start with the following identity:
\begin{eqnarray} 
& &(iT-\overline{P}_{<N>})e^{iT\overline{Q}_{<N>}}e^{iT\overline{R}_{<N>}}
\nonumber\\
=& &
iT (1- {\beta^{-1}_N} \sum_{\abs{k}\leq N} \tau_k(P) P_{k,N})
e^{iT\overline{Q}_{<N>}}e^{iT\overline{R}_{<N>}}
\nonumber\\
&- & {\beta^{-1/2}_N}\sum_{\abs{k}\leq N} \tau_k(P)
( 1- e^{-iT \overline{P}_{k,N}} -iT \overline{P}_{k,N} )
e^{iT\overline{Q}_{<N>}}e^{iT\overline{R}_{<N>}}
\nonumber\\
&-& {\beta^{-1/2}_N} \sum_{\abs{k}\leq N} \tau_k(P)
\rbk{  e^{-iT \overline{P}_{k,N}}- e^{-iT \overline{Q}_{k,N}}
e^{-iT \overline{R}_{k,N}} }
e^{iT\overline{Q}_{<N>}}
e^{iT\overline{R}_{<N>}}
\nonumber\\
&-&{\beta^{-1/2}_N} \sum_{\abs{k}\leq N} \tau_k(P)
e^{-iT \overline{Q}_{k,N}}
e^{-iT \overline{R}_{k,N}} 
e^{iT\overline{Q}_{<N>}}
e^{iT\overline{R}_{<N>}}
\nonumber\\
=& &
(A_1 + A_2  + A_3 ) e^{iT\overline{Q}_{<N>}}e^{iT\overline{R}_{<N>}}
+A_4 .
\label{eqn:d18}
\end{eqnarray}
Using the estimates of proof of Lemma 2.2 in \cite{XYCLT}  
(c.f. (2.8) and (2.10) of  \cite{XYCLT}) we obtain
$$\lim_{N \to \infty}  \varphi (A_a
  e^{iT\overline{Q}_{<N>}}e^{iT\overline{R}_{<N>}}) = 0$$
for $a=1,2$. 
On the other hand, 
\begin{eqnarray*}
 \norm{A_3} &\leq&
{\beta^{-1/2}_N} \sum_{\abs{k}\leq N} \norm{\tau_k(P)}
T^2 \norm{\sbk{\overline{Q}_{k,N} ,\overline{R}_{k,N}}}
\\
&\leq&
T^2  {\beta^{-3/2}_N} (2N+1) \norm{P} m(N)
\sum_{j \in \bfZ} \norm{\sbk{ \tau_j(Q) , R }}
\end{eqnarray*}
When we take $N$ to $ \infty$ we have
$ {\beta^{-3/2}_N} (2N+1)m(N) \to 0$. As a result
 $$\lim_{N \to \infty}  
\varphi (A_3 e^{iT\overline{Q}_{<N>}}e^{iT\overline{R}_{<N>}}) = 0.$$
\bigskip
\noindent
\par
We now look at $A_4$. We claim that
\begin{equation}
\lim_{N \to \infty} \beta^{1/2}_N
\norm{e^{-iT \overline{Q}_{k,N}}e^{-iT \overline{R}_{k,N}} 
e^{iT\overline{Q}_{<N>}}e^{iT\overline{R}_{<N>}}
- e^{iT( \overline{Q}_{<N>}-\overline{Q}_{k,N} )}
e^{iT( \overline{R}_{<N>}-\overline{R}_{k,N} )} } =0 .
\label{eqn:d19}
\end{equation}
To see this, we point out the following inequalities.
\begin{eqnarray}
& & \beta^{1/2}_N
\norm{\sbk{ e^{-iT \overline{R}_{k,N}}, e^{iT\overline{Q}_{<N>}}}}
\leq 2 T^2 \beta^{-1/2}_N m(N) \sum_{ j \in \bfZ}
\norm{ \sbk{ \tau_j(Q ) , R} }
\nonumber\\
& &\beta^{1/2}_N \norm{e^{-iT \overline{A}_{k,N}}
e^{iT\overline{A}_{<N>}}
- e^{iT( \overline{A}_{<N>}-\overline{A}_{k,N} )} }
\leq T^2 \beta^{-1/2}_N m(N) \sum_{ j \in \bfZ}
\norm{ \sbk{ \tau_j(A ) , A} } .
\label{eqn:d20}
\end{eqnarray}
for $A=P$ or  $Q$. Again for exponentially localized $A$,
$\sum_{ j \in \bfZ} \norm{ \sbk{ \tau_j(A ) , A} }$ is finite.
Thus  due to choice of $m(N)$  satisfying (\ref{eqn:d16}) , 
the right-hand side of (\ref{eqn:d20}) vanishes in the limit of 
taking $N$ to $\infty$.
\par
 The convergence (\ref{eqn:d19}) tell us that 
$$ -\lim_{N \to \infty} \varphi (A_4) = 
\lim_{N \to \infty}
\beta^{-1/2}_N \sum_{\abs{k}\leq N} 
\varphi ( \tau_k(P) e^{iT( \overline{Q}_{<N>}-\overline{Q}_{k,N}) }
e^{iT( \overline{R}_{<N>}-\overline{R}_{k,N}) } ) .$$
Now we concentrate on the following limit:
\begin{equation}
\lim_{N \to \infty} 
\varphi ( \tau_k(P) e^{iT( \overline{Q}_{<N>}-\overline{Q}_{k,N}) }
e^{iT( \overline{R}_{<N>}-\overline{R}_{k,N}) } ) .
\label{eqn:d21}
\end{equation}
We can estimate  decay of (\ref{eqn:d21}) as in \cite{XYCLT}.
The observable $\tau_k(P)$ is localized around the site $k$
with exponential tail while 
$ e^{iT( \overline{Q}_{<N>}-\overline{Q}_{k,N}) }
e^{iT( \overline{R}_{<N>}-\overline{R}_{k,N}) }$ 
is localized in $(-\infty , k-m(N)] \cup [k + m(N) , \infty)$
We have to approximate $Q$ and $R$ by strictly local
elements as in our proof of
CLT for non local observables of \cite{XYCLT}. 
(We omit the proof as the estimates is same as that of
\cite{XYCLT}.)
We arrive at the following estimate.
$$\abs{ \varphi ( \tau_k(P) e^{iT( \overline{Q}_{<N>}-\overline{Q}_{k,N}) }
e^{iT( \overline{R}_{<N>}-\overline{R}_{k,N}) } )}
\leq C\alpha ( m(N) )$$
This inequality shows that the limit (\ref{eqn:d21}) vanishes and
the claim of Lemma.
{\it End of Proof}
\bigskip
\noindent
\newline
Theorem 1.5 follows from Lemma 4.3 and Lemma 4.4.
\bigskip
\noindent
\bigskip
\noindent
\par
Next we consider the KMS condition. 
Theorem 1.7. follows easily from Theorem 1.5.
in the same manner as \cite{Goderis3}.
We repeat the argument briefly.
As the state $\tilde{\varphi}_{\beta}$ is quasifree,
it suffices to show KMS condition for
$\tilde{\varphi}_{\beta}(W(\tilde{Q}))$. 
 However as CLT is valid only for the selfadjoint part of $F_{\theta}$
we cannot consider  the complex time evolution
$\alpha_z$ in the central limit.
In fact , we can use another equivalent condition to the KMS condition.
(c.f. Proposition 5.3.6. of \cite{BratteliRobinsonII}.)
\begin{pro}
(i) If the state $\varphi$ is KMS, the time dependent correlation function 
$F(t) = \varphi(Q_1 \evl (Q_2))$
has a analytic extension to the strip
$I_{\beta} = \cbk{ z \vert 0 < Im z < \beta }$
$F(t) $ is bounded continuous on
 $\overline{I}_{\beta} = \cbk{ z \vert 0 \leq Im z \leq \beta }$.
and
\begin{equation}
F(t+i\beta ) = \varphi( \evl (Q_2) Q_1) . 
\label{eqn:d22}
\end{equation}
(ii)  Conversely if $F(t)$ admits a  analytic extension to
$I_{\beta}$ and  is bounded continuous on $\overline{I}_{\beta}$
satisfying (\ref{eqn:d22})for any $Q_1$ and $Q_2$
$\varphi$ is a KMS state.
\end{pro}
Furthermore we have the following bound for $F(z)$ on 
$\overline{I}_{\beta}$:
\begin{equation}
\abs{F(z)} \leq \norm{Q_1}\norm{Q_2} .
\label{eqn:d23}
\end{equation}
Let $\varphi_{\beta}$ be the unique Gibbs state of a finite range interaction. Set 
$$ F_N(t) = \varphi_{\beta} (e^{(i (Q_1)_{<N>}}e^{(i \evl (Q_2))_{<N>}} )$$
Due to (\ref{eqn:d23}), for any $N$ we have
$$\abs{ F_N(t)} \leq 1 .$$
Then we can choose a subsequence of $F_N(t)$ which converges
to a bound continuos function $F_{\infty}(t)$  on $\overline{I}_{\beta} $ and analytic on  $I_{\beta}$.
$F_{\infty}(t)$ satisfies the KMS condition on
$W (\tilde{F})$ which implies the same condition
on the von Neumann algebra generated by $\pi (W (\tilde{F}))$.

\section{Remarks}
\setcounter{theorem}{0}
\setcounter{equation}{0}
We include here one example of CLT
which was not discussed in \cite{XYCLT}.
We consider finitely correlated states of M.Fannes,B.Nachtergaele and 
R.Werner . (c.f. \cite{FNW} ) 
\par
The finitely correlated state is a non-commutative analogue of the 
(function of) Markov measure.
\par
Let $\varphi$ be a state  of
$\calA$.
Consider  the linear functional $\varphi_Q$ on 
$\calA_{(-\infty , k]}$ defined by
\begin{equation}
\varphi_Q(R)= \varphi(QR)  \quad \quad 
\mbox{ for $Q$ in $\calA_{[k+1,\infty)}$ .}
\label{eqn:e1}
\end{equation}
Set 
\begin{equation}
 {\cal S}_k (\varphi) = 
\cbk{ \varphi_Q \; \vert \;  Q \in \calA_{[k+1,\infty)}  } . 
\label{eqn:e2}
\end{equation}
\begin{df}
$\varphi$ is a finitely correlated state 
if $ {\cal S}_k $ defined in (\ref{eqn:e2}) is  a finite dimensional
 subspace of the set of all linear functionals on   
$\calA_{(-\infty , k]}$ for any k.
\end{df}
\begin{pro}
Let $\varphi$ be  a translationally invariant finitely correlated state.
Suppose that  $\varphi$ is mixing in the following sense: 
$$\lim_{k \to  \infty} \varphi (Q_1 \tau_k(Q_2)) = 
\varphi(Q_1)\varphi(Q_2).$$
Then, the uniform exponential mixing (\ref{eqn:a11}) holds and
the central limit theorem is valid.
\end{pro}
Note that the exponential decay of two point correlation
$$\abs{ \varphi (Q_1 \tau_k(Q_2)) - \varphi(Q_1)\varphi(Q_2)}
\leq C(Q_1 ,Q_2) e^{-Mk} $$
is known. What matters here is the constant $C(A,B)$ on $Q_1$ 
and $Q_2$.
\bigskip
\noindent
\newline
{\it Proof}:
To prove the above proposition,  we consider
a translationally invariant state $\varphi$ of $\calA$
and the GNS triple 
$\cbk{ \pi_{ \varphi}(\calA) , \calH_{ \varphi} , \Omega_{ \varphi} }$ 
associated with $\varphi$.
Let $U$ be the unitary on $\calH_{ \varphi}$ implementing the shift 
$\tau_1$.
$$ U \pi_{ \varphi}(Q) U^* = \pi_{ \varphi}(\tau_1 (Q))
\quad \quad U \Omega_{ \varphi}= \Omega_{ \varphi}$$
Let  $P$ be the projection with the following range :
$$\calH_{(-\infty ,-1]} =
\overline{\pi_{ \varphi}(\calA_{(-\infty ,-1]})\Omega_{ \varphi}}.$$ 
As the unitary $U^*$ leaves $\calH_{(-\infty ,-1]}$ invariant, we have
$$PU^*P = U^*P  . $$
Set 
$$\calN = P\pi_{ \varphi}(\calA_{[0,\infty )})P .$$ 
As $\tau_1$ is implemented by a unitary $U$ on 
$\calH_{ \varphi}$, $\tau_1$ is extended to $\calN$ via the formula
$$ \tau_1(PQP) = PUQU^*P .$$
We define the completely positive unital map $L_{dual}$
from $\calN$ to $\calN$
via the following equation: 
$$L_{dual}(R )  = PU R U^*P .$$
where $R$ is an element of $\calN$. 
\par
We call $L_{dual}$ the dual transfer operator.
Consider the vector state $\psi_{dual}$ 
of $\calN$ associated with  $\Omega_{ \varphi}$.
$\psi_{dual}$ is faithful as 
$\calN$ commutes with $\calA_{(-\infty ,-1]}$
and $\Omega_{ \varphi}$ is a cyclic vector for
$\calA_{(-\infty ,-1]}$ in $\calH_{(-\infty ,-1]}$.
\begin{lmm}
The dimension of $\calN$ is finite if and only if
the state $\varphi$ is finitely correlated.
\end{lmm}
{\it Proof}:
Set
$$\calK = P\pi_{ \varphi}(\calA_{[0, \infty )})\Omega_{ \varphi}.$$
Note that
$$\varphi_Q(R)=\rbk{P\pi_{ \varphi}(Q^*)\Omega_{ \varphi} ,
\pi_{ \varphi}(R)\Omega_{ \varphi}} .$$
By the Schwartz inequality $\varphi_Q$ is a bounded linear functional
on the Hilbert space $\calH_{(-\infty ,-1]}$ and due to Riesz Lemma
the set $\calS_0$ is isomorphic to the subspace $\calK$ of 
$\calH_{(-\infty ,-1]}$
To see the above Lemma, it suffices to notice that the dimension of
$\calK$ is finite if and only if $\varphi$ is finitely correlated.
{\it End of Proof}
\bigskip
\noindent
\bigskip
\noindent
\par
Due to Lemma 4.3, for a finitely correlated $\varphi$, the dual transfer operator $L_{dual}$ is a linear operator on the finite dimensional space 
$\calN$. In fact we can extend the dual transfer operator
$L_{dual}$ to a unital completely positive map on the set of 
bounded linear operators on  $\calK$.
The norm  of $L_{dual}$ is one, $\norm{L_{dual}}=1$.
Furthermore, due to the mixing property of $\varphi$
$$ \lim_{N \to \infty} L_{dual}^n (A) = \psi_{dual}(A) 1$$ 
The eigenvalue $1$ of the dual transfer operator $L_{dual}$ 
is non-degenerate and there exists no other peripheral spectrum due to mixing of the state $\varphi$.  
Other eigenvalues of $L_{dual}$ is strictly less than one.
(c.f. \cite{FNW} )
Thus there exist constants $C$ and $M$ such that
$$ \norm{L_{dual}^n (A) -\psi_{dual}(A) 1} \leq C e^{-Mn} \norm{A} .  $$
Then for Q in $\calA_{(-\infty , -1]}$ , R in $\calA_{[0, \infty)}$
and $n$ positive,  
\begin{eqnarray}
 \abs{\varphi(A\tau_n(B) ) -\varphi(A)\varphi(B) } &\leq&
\norm{L_{dual}^n (P\pi_{\varphi}(B)P) -
\psi_{dual}(P\pi_{\varphi}(B)P) 1} \norm{A}
\nonumber\\
&\leq& C e^{-Mn} \norm{A}\norm{B}
\label{eqn:e3}
\end{eqnarray}
{\em End of Proof of Proposition 4.2.}
\bigskip
\noindent
\bigskip
\noindent
\par
In Section 1 we considered Gibbs states of finite range interactions
for one-dimensional chains.
The formalism can be extended to long range interactions. However,
we are unable to prove necessary mixing condition for CLT.
To clarify the difficulty in our approach we present the results
we can prove by our methods.  
\par
First we replace $F_{\theta}$ with suitable non-commutative H{\"o}lder
continuos functions.
In stead of the weight $\theta^n$ we take $n^{-\eta}$ with the condition
$\eta > 3$ and set
$$\calF^{(\eta)} = \cbk{ Q \in \calA \:\: \vert \:\: \sum_{n=0}^{\infty} 
\norm{Q}^{(n)}  n^{\eta} < \infty }$$
\begin{ass}
The interaction $\psi$ is translationally
invariant and we assume two conditions
\newline
(i) The time evolution is well-defined on $\calA$.
$$ \lim_{N \to \infty}  e^{it H_{[-N,N]}} Q e^{-it H_{[-N,N]}}
=\evl (Q) $$
(ii) The following limit exists and it gives rise to an element of 
$\calF^{(\eta)} $.
\begin{equation}
\lim_{N \to \infty}  e^{\beta/2 H_{[0, N]}} e^{-\beta/2 H_{[1,N]}}
= a_{\beta}  \in \calF^{(\eta)}
\label{eqn:e4}
\end{equation}
\end{ass}
\begin{pro}
If  Assumption 4.4 is valid, the Gibbs state for the inverse temperature
$\beta$ is unique, satisfying the following mixing:
\begin{equation}
\abs{\varphi(Q_1 \tau_n(Q_2)) -
\varphi(Q_1 )\varphi(Q_2)} \leq K n^{-(\eta -1)}
\norm{Q_1}\norm{Q_2}
\label{eqn:e5}
\end{equation}
for  $Q_1$ in ${\calA}_{(-\infty, -1]}$, $Q_2$ in ${\calA}_{[0,\infty)}$ 
and $n >0$.
\par
Thus as a corollary, CLT holds if $\eta >3$.
\end{pro}
Decay of correlation (\ref{eqn:e5}) can be derived by analysis
 of Ruelle transfer operators (c.f. \cite{TM2} ), however we have no idea 
to prove the convergence of (\ref{eqn:e4})
\bigskip
\noindent
\bigskip
\noindent
\par
So far, we have discussed quantum spin models on one-dimensional lattice only.  In higher dimensional lattice we can prove
CLT if we have the following mixing for a state $\varphi$
$$  \abs{\varphi (Q_1 Q_2) - \varphi(Q_1)\varphi(Q_2)}
\leq C_{\Lambda_0} e^{- M k} \norm{Q_1}  \norm{Q_2} $$
where the support of $Q_1$ is $\Lambda_0$ and the distance of
the support of $Q_1$ and that  of $Q_2$ is $k$ and the constant
$ C_{\Lambda_0}$ is independent of the size of the support of $Q_2$.
We are not certain that this estimate ( or the (in)dependence of 
the constant $ C_{\Lambda_0}$ on the size of the support of $Q_2$ )
is valid for high temperature Gibbs states at the moment.
It is a non trivial question whether the above estimate can be shown by
the high temperature expansion. 
Note that we do not require the assumption (CLT4) of \cite{Goderis2}.


\end{document}